\documentclass[a4paper,12pt]{article}
\usepackage{graphicx}
\usepackage{color}
\usepackage{placeins}
\usepackage{float}
\usepackage{hyperref}
\usepackage{tabularx,colortbl}
\usepackage[T1]{fontenc}
\usepackage[utf8]{inputenc}
\usepackage{amsfonts}
\usepackage{amssymb}
\usepackage{color}
\usepackage{mathptmx,textcomp}
\usepackage{amsmath,mathrsfs,bm,cases}
\usepackage[labelfont=small,textfont=small,labelsep=period,justification=centering]{caption,subfig}
\usepackage{siunitx}
\usepackage{enumitem}
\usepackage{multirow}

\usepackage{amsbsy}
\usepackage{amsmath}
\usepackage{amsfonts}
\usepackage{amssymb}

\def\bE{\ensuremath{\mathbf{E}}}

\newcommand{\Lvect}{\mathscr{M}}

\def\curl{\mathbf{curl}\,}

\def\exp{\mathrm{exp}}

\def\tensmu{\boldsymbol{\mu}_r}
\def\tenseps{\boldsymbol{\varepsilon}_r}
\def\tensepsb{\boldsymbol{\varepsilon}_{r,b}}
\def\tensmub{\boldsymbol{\mu}_{r,b}}

\newcommand{\tensid}{\boldsymbol{\textrm{I}}}
\newcommand{\tensepss}{\boldsymbol{\varepsilon}_{r,s}}
\newcommand{\tensmus}{\boldsymbol{\mu}_{r,s}}
\newcommand{\Mnm}{\mathbf{M}_{n,m}}
\newcommand{\Nnm}{\mathbf{N}_{n,m}}
\newcommand{\FMnm}{\mathbf{F}^{d,(M)}_{n,m}}
\newcommand{\FNnm}{\mathbf{F}^{d,(N)}_{n,m}}
\newcommand{\RE}{\Re {\it{e}}}
\newcommand{\IM}{\Im {\it{m}}}
\definecolor{Comments}{rgb}{0.9,0.05,0.05}
\definecolor{Comments2}{rgb}{0.4,0.04,0.05}
\definecolor{Check}{rgb}{0.9,0.5,0.5}
\definecolor{black}{rgb}{0,0,0}

\definecolor{colorrewier1}{RGB}{66, 134, 244}
\definecolor{colorrewier2}{RGB}{244, 110, 66}
\def\crevi{\color{black}}
\def\crevii{\color{black}}

\author{Guillaume Demésy$^{1,}$\footnote{Corresponding author : \texttt{guillaume.demesy@fresnel.fr}} , Brian Stout$^1$ and Jean-Claude Auger$^2$\\
\footnotesize{$^1$ Aix-Marseille Université, CNRS, Centrale Marseille, Institut Fresnel UMR 7249, 13013 Marseille, France.}\\
\footnotesize{$^2$ Kyolaris-Research Ltd, 2003, 20/F, Tower 5, China Hong Kong City,} \\
\footnotesize{33 Canton Road, Tsim sha Tsui, Kowloon, Hong Kong.}
}

\title{Scattering matrix of arbitrarily shaped objects: Combining Finite Elements and Vector Partial Waves.}

\begin{document}
  \maketitle

\begin{abstract}
	We demonstrate the interest of combining Finite Element calculations with the Vector 
	Partial Wave formulation (used in $T$-matrix and Mie theory) in 
	order to characterize the electromagnetic scattering properties of isolated individual scatterers. 
	This method consists of individually feeding the finite element problem with incident Vector Partial 
  Waves in order to numerically determine the $T$-matrix elements of the scatterer.
  For a sphere and an spheroid, we demonstrate that this method determines the
  scattering matrix to high accuracy. 
   Recurrence relations for a fast determination of the vector partial waves are given explicitly,
   and an open-source code allowing the retrieval of the presented numerical results is provided.
\end{abstract}
%
%
%

\section{Introduction} 
Full 3D time-harmonic computations of vector fields of wavelength $\lambda$, interacting with a collection of
objects of arbitrary shapes and sizes inside volumes of a few thousand $\lambda^3$, remains a formidable numerical challenge. At this
scale, resonant processes are non-negligible and geometric optics is not yet applicable. When attempting to fill this void, it is
common to combine the strengths of two different numerical techniques. For instance, in the case of periodic structures, Finite Element
Methods (FEM, based on a space discretization) and Fourier Modal Method (FMM, based on a Fourier reciprocal space discretization) have been
combined in order to benefit from the speed of the FMM when dealing with thick layers and straight walls together with the versatility
of the FEM with respect to opto-geometric parameters of diffractive elements \cite{dossou2012modal,hu2014convergence,huber2009simulation}.

The idea developed in this paper is to combine FEM calculations with the multipolar Vector Partial Wave (VPW) formalism used in the $T$-matrix
theory. {\crevi Note that Mie theory refers to spherical scattering particles, we denote here by ``$T$-matrix formalism'' the theory adapted to
non-spherical particles introduced by Waterman \cite{waterman1971symmetry} and extended by Mishchenko \cite{mishcheko1991}.}
The VPW formalism has the advantage that it allows the multiple-particle problem to be treated very accurately and fast, but it
requires knowledge of the $T$-matrix of the individual scatterers. The $T$-matrix in the VPW framework can be viewed as a complete solution
to the scattering problem for an arbitrary incident field. Although there is a vast literature on methods to calculate the $T$-matrix,
most techniques require the the properties of the scatterers to obey certain assumptions, like single valued surface functions,
piecewise continuity, axial symmetry, and so forth \cite{mishcheko2010}.

We show how the FEM can be used to directly determine the multipolar $T$-matrix elements of particles of arbitrary shape and/or constitutive
material(s). {\crevii A first advantage is that the field scattered by a single arbitrary object illuminated by an arbitrary source is determined
everywhere outside the circumscribing sphere of the object thanks to the $T$-Matrix formalism. Another advantage lies in the fact that
once the full $T$-matrix of the individual particles are determined, extremely efficient multipolar, multi-scattering methods are
applicable to 3D collections of particles with arbitrarily large inter-particle and possibly non-periodic separations
\cite{auger2007mie,stout2008}. In the opposite situation of small inter-particle distances, the superposition $T$-matrix method is not
applicable to systems of non-spherical particles once a particle intersects an adjacent particle's circumscribing sphere
\cite{auger2007mie}. The presented method allows to partly circumvent this issue by calculating the $T$-matrix of the assembly of the
two (or more) close particles with individual interpenetrating circumscribing spheres. This requires to consider the larger
circumscribing sphere encompassing the two (or more) objects. However, our approach does not fully solve the problem when the
quantity of interest is the near field in-between the two particles since only the field outside the larger circumscribing sphere
can be computed from the $T$-matrix.}

In this manner we make use of the respective strengths of two quite different approaches commonly used in rigorous treatments of the vector
harmonic Maxwell equations in open scattering electromagnetic problems: The versatility of the FEM with the large-scale multiple
scattering strengths of multipolar $T$-matrix calculations. Upon completion of this manuscript, it was brought to our attention that a similar
approach has been published very recently \cite{fruhnert2017computing}. Nevertheless, these authors privileged a multiple incident plane waves
approach, whereas we illuminate the structure directly with stationary multipolar harmonic fields.

Finite Element Methods (FEM) represent a very general set of techniques for determining approximate solutions of partial
derivative equations, like Helmholtz-type propagation studied herein. Their main advantage lies in their ability to handle arbitrary
geometries through unstructured volume meshes of the domain of interest: The discretization of oblic geometry edges and graded indexed
materials (\emph{e.g.} metamaterials) are naturally built in. The key ingredients to model the general 3D electromagnetic scattering
problem using FEM are (i) appropriate basis functions allowing field discontinuities (Whitney elements), (ii) an unknown field
satisfying a proper outgoing wave condition, and (iii) a way to bound the infinite background medium -- Perfectly Matched Layers (PML)
\cite{teixeira1997systematic} have proven to be very effective tools to that effect. A good choice is usually to calculate a diffracted
field rather than the total field since it allows to bring the sources of the incident electromagnetic radiation such as plane waves
within the diffractive elements as detailed in \cite{demesy2009ol}. Contrary to Fourier methods, the raw result given by the FEM is a
3D vector field map around a close vicinity of the diffractive element. Energy related quantities of common interest, such as the
scattering and extinction efficiencies, can be post-processed using classical Fourier-Bessel expansions of spherical cuts of the field
around the scatterer. Absorption efficiency can be obtained by integration of the square norm of the electric field in lossy regions.

In spite of constant advances made in the field of linear algebra for sparse matrices \cite{petsc-web-page} and a steadily increasing computing
power, the main drawback of Finite Elements remains the large amount of memory required by fast direct solvers. This is a direct
consequence of the high degree of connectivity of the unknowns in 3D, especially for high order schemes. Let us mention that the Domain
Decomposition (DDM) \cite{benamou1997domain,voznyuk2013scattered} is a promising technique to tackle large domains by splitting the whole
problem into smaller sub-problem governed by a global interface (surface) problem. Finally, let us mention that a combination of FEM and
$T$-matrix theory is usually implicitly invoked when computing the radiation pattern \cite{Bohr83,stout01,jin2014finite,felbacq1994scattering},
allowed by the asymptotic of form of spherical Hankel functions as $|kr|\rightarrow+\infty$.

The motivations of this work are three fold:
(i)	 Translation/Addition theorems \cite{stout2002transfer} allow to combine elementary 
scattering matrices of independent scatterers to establish a global scattering 
matrix of the assembly of scatterers in a particular geometrical configuration.
(ii) The knowledge of the scattering matrix allows the fast computation 
of the diffracted field by any source, which is extremely valuable for fast 
evaluation of parametric studies, such as the angular response of non-spherical particles, 
where pure FEM evaluation would require a new large sparse matrix inversion for every new source.
(iii) Physical interpretations of multipolar developments, related to the  
notion of modes, gives a more in-depth insight than a short-sighted 3D field map.

In this paper, we propose to consider an arbitrarily shaped object and to use the FEM
to retrieve its scattering matrix expressed in the outgoing vector partial waves basis
traditionally used in the Mie Theory and T-matrix formalism, for the purpose of embedding this
as an elementary brick in existing multiple scattering schemes. The paper is
organized as follows. We first present the Finite Element formulation of the
scattering problem. Contrary to recent propositions in the literature
\cite{varault2013multipolar}, we propose to directly input the Finite Element model
with the appropriate vector partial waves. Next, the vector partial waves expansion is
described. All relevant recurrence relations for fast evaluation of spherical
functions involved in the expansion are given in the Appendix. The numerical validity of
the method is demonstrated by comparing to Mie theory results for a sphere and a Waterman method
calculation of a spheroid. {\crevi An open-source model (based on Onelab, Gmsh \cite{gmsh} and GetDP
\cite{getdp}) allowing to retrieve the numerical results is provided \cite{code}.}

\begin{figure}[h]
		\begin{center}
      \includegraphics[width=.82\columnwidth]{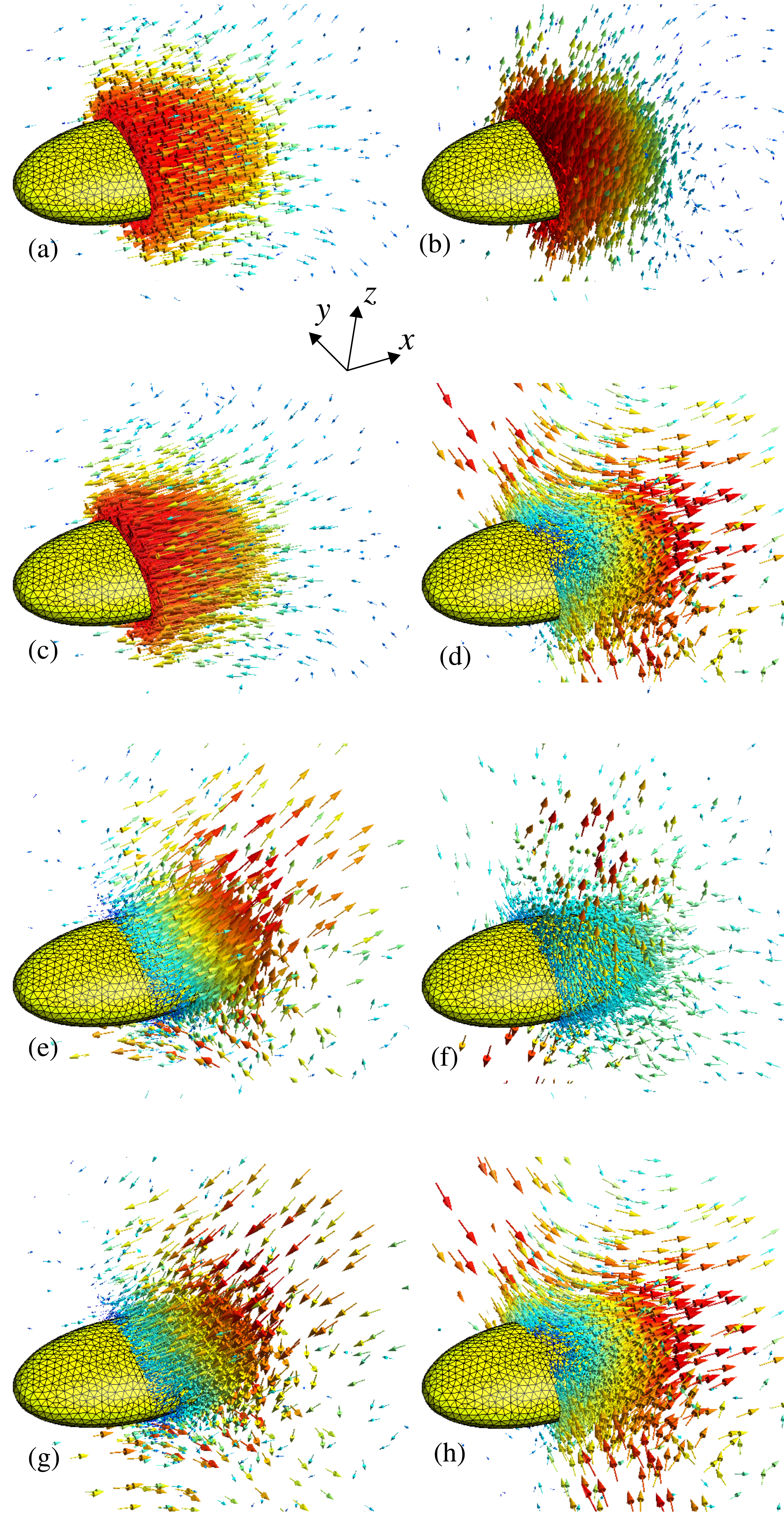}
		\caption{
    (a)~$\RE\{\mathbf{F}^{d,(N)}_{1,-1}\}$. (b)~$\RE\{\mathbf{F}^{d,(N)}_{1,0 }\}$. 
    (c)~$\RE\{\mathbf{F}^{d,(N)}_{1,1 }\}$. (d)~$\RE\{\mathbf{F}^{d,(N)}_{2,-2}\}$. 
    (e)~$\RE\{\mathbf{F}^{d,(N)}_{2,-1}\}$. (f)~$\RE\{\mathbf{F}^{d,(N)}_{2,0 }\}$. 
    (g)~$\RE\{\mathbf{F}^{d,(N)}_{2,1 }\}$. (h)~$\RE\{\mathbf{F}^{d,(N)}_{2,2 }\}$.
    Real part of the total fields $\FNnm$ for $n_{\mathrm{max}}=2$ computed using the FEM in the case 
    of an spheroid. For clarity, the fields are shown for $x>0$ only. 
    The surface mesh of the spheroid is shown for $x<0$.
    } \label{fig:realFNnm}
		\end{center}
\end{figure}

\section{Finite Element formulation} 
Consider a scatterer of relative permittivity and permeability tensors fields denoted
$\tensepss(\mathbf{x})$ and $\tensmus(\mathbf{x})$ lying inside an isotropic and homogeneous background of constant permittivity
and permeability tensors $\tensepsb =\varepsilon_b\,\tensid$ and $\tensmub =\mu_b\,\tensid$. The resulting relative permittivity
and permeability tensors fields $\tenseps(\mathbf{x})$ and $\tensmu(\mathbf{x})$ of the scattering problem, defined over
$\mathbb{R}^3$, are respectively (resp.) equal to $\tensepss(\mathbf{x})$ and $\tensmus(\mathbf{x})$ inside the scatterer, and
$\varepsilon_b\,\tensid$ and $\mu_b\,\tensid$ outside.

Our approach can be applied irrespective of whether $\tensmus$ and $\tensepss$ tensors 
fields are full with possibly complex component functions of $\textbf{x}$.
They describe a possibly fully anisotropic, lossy or active, graded-index scatterer.

One the one hand, the total electric field $\bE$ solution of a diffraction problem 
involving any distant Maxwellian source satisfies the vector Helmholtz propagation 
equation in the time-harmonic regime:
\begin{equation}\label{eq:defL}
		\Lvect_{\tenseps,\tensmu} (\bE):=-\curl\left(\tensmu^{-1}\,\curl\bE\right) + k_0^2\,\tenseps\,\bE =
		\textbf{0}\,.
\end{equation}

\sisetup{round-mode=places, round-precision=4}
\begin{table*}[ht]
	\begin{center}
		\begin{tabular}{ll c c c c c c c}
			\hline
		Source										& Expansion								 & $T_{i,j}$	&	 $\RE\{T_{i,j}\}$						& $\IM\{T_{i,j}\}$						& $|2\,T_{i,j}+1|$	& relative   & $\sigma^{\mathrm{cuts}}_{i,j}$	 \\
		type										& coefficient								&	 &	& &  &	error/Mie	\\ \hline
		$ \mathbf{M}_{1,1}^{(1)}$ & $ f_{1,1}^{(h)}$					 & $T_{1,1}$	&	 \num{-1.2385259697e-02}		&	 \num{-1.1056579997e-01}		& 0.999986			&	\num[round-mode=places, round-precision=1]{1.1e-03}	 & \num[round-mode=places, round-precision=1]{3.258e-06} \\
		$ \mathbf{M}_{1,0}^{(1)}$ & $ f_{1,0}^{(h)}$					 & $T_{2,2}$	&	 \num{-1.2385169286e-02}		&	 \num{-1.1057000518e-01}		& 0.999988			&	\num[round-mode=places, round-precision=1]{1.1e-03}	 & \num[round-mode=places, round-precision=1]{3.045e-06} \\
		$\mathbf{M}_{1,-1}^{(1)}$ & $f_{1,-1}^{(h)}$					 & $T_{3,3}$	&	 \num{-1.2385200511e-02}		&	 \num{-1.1056590287e-01}		& 0.999986			&	\num[round-mode=places, round-precision=1]{1.1e-03}	 & \num[round-mode=places, round-precision=1]{3.694e-06} \\ \hline
										\multicolumn{3}{c }{Mie}													  &	 \num{-1.24056e-02     }    &	 \num{-1.10688e-01		 }		&										 \multicolumn{3}{c}{}																		\\ \hline
		$ \mathbf{N}_{1,1}^{(1)}$ & $ f_{1,1}^{(e,\parallel)}$ & $T_{4,4}$	&	 \num{-8.1416217938e-02}		&	 \num{-2.7344033443e-01}		& 0.999964			&	\num[round-mode=places, round-precision=1]{4.1e-04}	 & \num[round-mode=places, round-precision=1]{2.707e-05} \\
		$ \mathbf{N}_{1,0}^{(1)}$ & $ f_{1,0}^{(e,\parallel)}$ & $T_{5,5}$	&	 \num{-8.1414532024e-02}		&	 \num{-2.7344674202e-01}		& 0.999974			&	\num[round-mode=places, round-precision=1]{4.0e-04}	 & \num[round-mode=places, round-precision=1]{3.259e-05} \\
		$\mathbf{N}_{1,-1}^{(1)}$ & $f_{1,-1}^{(e,\parallel)}$ & $T_{6,6}$	&	 \num{-8.1415303240e-02}		&	 \num{-2.7344188994e-01}		& 0.999967			&	\num[round-mode=places, round-precision=1]{4.1e-04}	 & \num[round-mode=places, round-precision=1]{2.776e-05} \\ \hline
		$ \mathbf{N}_{1,1}^{(1)}$ & $ f_{1,1}^{(e,\perp)}$		 & $T_{4,4}$	&	 \num{-8.1384568157e-02}		&	 \num{-2.7343965268e-01}		& 1.000016			&	\num[round-mode=places, round-precision=1]{4.7e-04}	 & \num[round-mode=places, round-precision=1]{2.994e-04} \\
		$ \mathbf{N}_{1,0}^{(1)}$ & $ f_{1,0}^{(e,\perp)}$		 & $T_{5,5}$	&	 \num{-8.1368265239e-02}		&	 \num{-2.7342113303e-01}		& 1.000023			&	\num[round-mode=places, round-precision=1]{5.6e-04}	 & \num[round-mode=places, round-precision=1]{5.510e-04} \\
		$\mathbf{N}_{1,-1}^{(1)}$ & $f_{1,-1}^{(e,\perp)}$		 & $T_{6,6}$	&	 \num{-8.1387379353e-02}		&	 \num{-2.7344185603e-01}		& 1.000014			&	\num[round-mode=places, round-precision=1]{4.6e-04}	 & \num[round-mode=places, round-precision=1]{3.072e-04} \\ \hline
										\multicolumn{3}{c }{Mie}													  &	 \num{-8.14651e-02     }    &	 \num{-2.73548e-01		 }		&										 \multicolumn{3}{c}{}																		\\ \hline
		\end{tabular}
	\end{center}
	\centering \caption{Theoretically non zero $T$-matrix (\emph{i.e.} diagonal) coefficients computed using the FEM for a sphere of diameter $\lambda/4$ with $\tensepss=9\,\tensid$.}
  \label{tab:sphere_T}
\end{table*}

Note that the choice of the electric field formulation is arbitrary and 
that the electromagnetic source of the problem has yet to be defined.
One the other hand, vector partial waves $\Mnm^{(1)}$ and $\Nnm^{(1)}$ 
defined in Eq.~(\ref{RgMetN}), are stationary solutions of the vector 
Helmholtz equation in a homogeneous space made of the background characteristics:
\begin{equation}\label{eq:helm_MN}
	\Lvect_{\tensepsb,\tensmub}(\Mnm^{(1)})=\Lvect_{\tensepsb,\tensmub}(\Nnm^{(1)})=\textbf{0}\,.
\end{equation}

Out of the linearity of the operator $\Lvect$, the underlying elementary diffraction problem
amounts to \cite{demesy2010all} look for diffracted fields $\FMnm:=\bE-\Mnm^{(1)}$ and $\FNnm:=\bE-\Nnm^{(1)}$ such that:
\begin{equation}\label{eq:helmdif_M}
	\begin{array}{ll}
		\Lvect_{\tenseps,\tensmu}(\FMnm)=& 
		- \curl\left[ \left(\tensmub^{-1}-\tensmu^{-1}\right)\,\curl\, \Mnm^{(1)} \right]\\
				&+k_0^2\,(\tensepsb-\tenseps)\,\Mnm^{(1)} 
	\end{array}
\end{equation}
and
\begin{equation}\label{eq:helmdif_N}
	\begin{array}{ll}
		\Lvect_{\tenseps,\tensmu}(\FNnm)=& 
		- \curl\left[ \left(\tensmub^{-1}-\tensmu^{-1}\right)\,\curl\, \Nnm^{(1)} \right]\\
		 &+k_0^2\,(\tensepsb-\tenseps)\,\Nnm^{(1)} \, ,
	\end{array}
\end{equation}
where both $\FMnm$ and $\FNnm$ satisfy an outgoing wave condition. In Eqs.~(\ref{eq:helmdif_M}-\ref{eq:helmdif_N}), 
the right hand side terms turn out to be known source terms since they only involve the working frequency, 
the contrast of electromagnetic properties between the scatterer and the background, and the 
incident field. Unknowns fields $\FNnm$ and $\FMnm$ correspond to electromagnetic fields radiated from
the scatterer. This allows to safely truncate the computational domain using PMLs \cite{Lassas2}. 
In practice, 3D geometries and conformal tetrahedral mesh are computed using the 
Gmsh GNU software \cite{gmsh} and the discretization of the weak equation associated with 
Eqs.~(\ref{eq:helmdif_M}-\ref{eq:helmdif_N}) is performed thanks to the flexibility of the 
finite element GNU software GetDP \cite{getdp}. Some examples of computed $\FNnm$ fields are shown 
in Fig.~\ref{fig:realFNnm}. A self-consistent model based on the Onelab interface is given in Ref.~\cite{code}

\section{Vector partial wave expansion}
Any scattered or outgoing field, such as $\FMnm$ (resp.~$\FNnm$) solution of 
Eq.~(\ref{eq:helmdif_M}) (resp.~Eq.~(\ref{eq:helmdif_N})), can be expanded as:
\begin{equation}
\mathbf{F}^{d,(M,N)}_{n,m}\left(	\mathbf{r}\right)	 =\sum_{n=1}^{n_{\max}
}\sum_{m=-n}^{m=n}\left[	\mathbf{M}_{nm}^{\left(	 +\right)	 }(k\mathbf{r}
)f_{nm}^{\left(	 h\right)	 }+\,\mathbf{N}_{nm}^{\left(	+\right)
}(k\mathbf{r})f_{nm}^{\left(	e\right)	}\right]\,\,,
\end{equation}
where the outgoing vector partial waves $\mathbf{M}_{nm}^{\left(+\right)}$ 
and $\mathbf{N}_{nm}^{\left(+\right)}$  are defined in Section~\ref{sec:pwe} (see Eq.~(\ref{eq:defMN})).
{\crevi Note that a time dependence in $e^{+i\omega t}$ is assumed throughout the paper}.
Recurrence relations used in our code for their fast numerical evaluation are self-consistently detailed.

The so-called {\crevi transition matrix $T$} relates the diffracted field (\emph{i.e.}
the coefficients of its expansion over $\Mnm^{(+)}$ and $\Nnm^{(+)}$ waves) 
to an incident stationary vector partial wave $\Mnm^{(1)}$ or $\Nnm^{(1)}$.
Note that the double indices $n,m$ are conveniently 
combined into a single integer $p$ according to the bijective relations detailed in
Sec.~\ref{sec:indexing}.
In this manner the $T$-matrix is described as a square block matrix  of dimensions $2p_{max}\times 2p_{max}$
constituted of four $T^{i,j}$ submatrices $p_{max}\times p_{max}$ according to the following convention:
\begin{equation}
	T=
		\begin{bmatrix}
			T^{1,1} & T^{1,2}\\
			T^{2,1} & T^{2,2}
		\end{bmatrix}\, ,
\end{equation}
where each coefficient of $T$ correspond to:
\begin{itemize}
	\item $T^{1,1}_{p,p'}=f^{(h)}_{p'}$ for a source of type $\mathbf{M}^{(1)}_{p}$,
	\item $T^{2,1}_{p,p'}=f^{(e)}_{p'}$ for a source of type $\mathbf{M}^{(1)}_{p}$,
	\item $T^{1,2}_{p,p'}=f^{(h)}_{p'}$ for a source of type $\mathbf{N}^{(1)}_{p}$ and
	\item $T^{2,2}_{p,p'}=f^{(e)}_{p'}$ for a source of type $\mathbf{N}^{(1)}_{p}$.
\end{itemize}

In practice, the computation of a full matrix $T$ truncated at $p_{max}$ involves the FEM computation of $2\times p_{max}$ sub-problems:
(i) Setting one of the $\Mnm^{(1)}$ (resp. $\Nnm^{(1)}$) as a source, (ii) computing the direct problem associated with
Eq.~(\ref{eq:helmdif_M}) (resp.~Eq.~(\ref{eq:helmdif_N})) using the FEM, and (iii) post-processing by classical numerical integration
the $p_{max}$ expansion coefficients of the resulting scattered field $\FMnm$ (resp. $\FNnm$) over the $\Mnm^{(+)}$ (resp. $\Nnm^{(+)}$)
outgoing vector partial waves. In fact, each of these sub-problem involves one FEM run and leads to the determination one full column of
the $T$-matrix.

\section{Results for a sphere} 
We consider a dielectric sphere of diameter $\lambda/4$ with relative permittivity $\varepsilon_r=9$
in a homogeneous background of relative permittivity of 1. As for the numerical parameters involved, cartesian PMLs of
thicknesses set to $\lambda$ are used, the distance between the sphere and the PMLs is set to $\lambda/4$.

The edges of the tetrahedral mesh have a characteristic size of $\lambda/(\sqrt{\varepsilon_r}N)$ in free space and PML regions, where
the value of $N$ controls the number of mesh elements per wavelength inside the considered material. Note that all materials are
transparent in this example. {\crevi When dealing with lossy materials as metals, the relevant length scale is no longer the
apparent wavelength inside the material, but the skin depth, so the mesh size is set accordingly.} Second order shape
functions are used. Several radial cuts of the diffracted fields are performed (10 cuts ranging from $R=r_{\mathrm{sph}}+\lambda/40$ to
$R=r_{\mathrm{PML}}-\lambda/40$) in the background to extract the $T$-matrix coefficients, as detailed above. Theoretically, the
$T$-matrix coefficients do not depend on the cut radius, but, as shown later on, taking several cuts
presents many advantages for probing the numerical precision.

\begin{figure}[h]
		\begin{center}
      \includegraphics[width=\columnwidth]{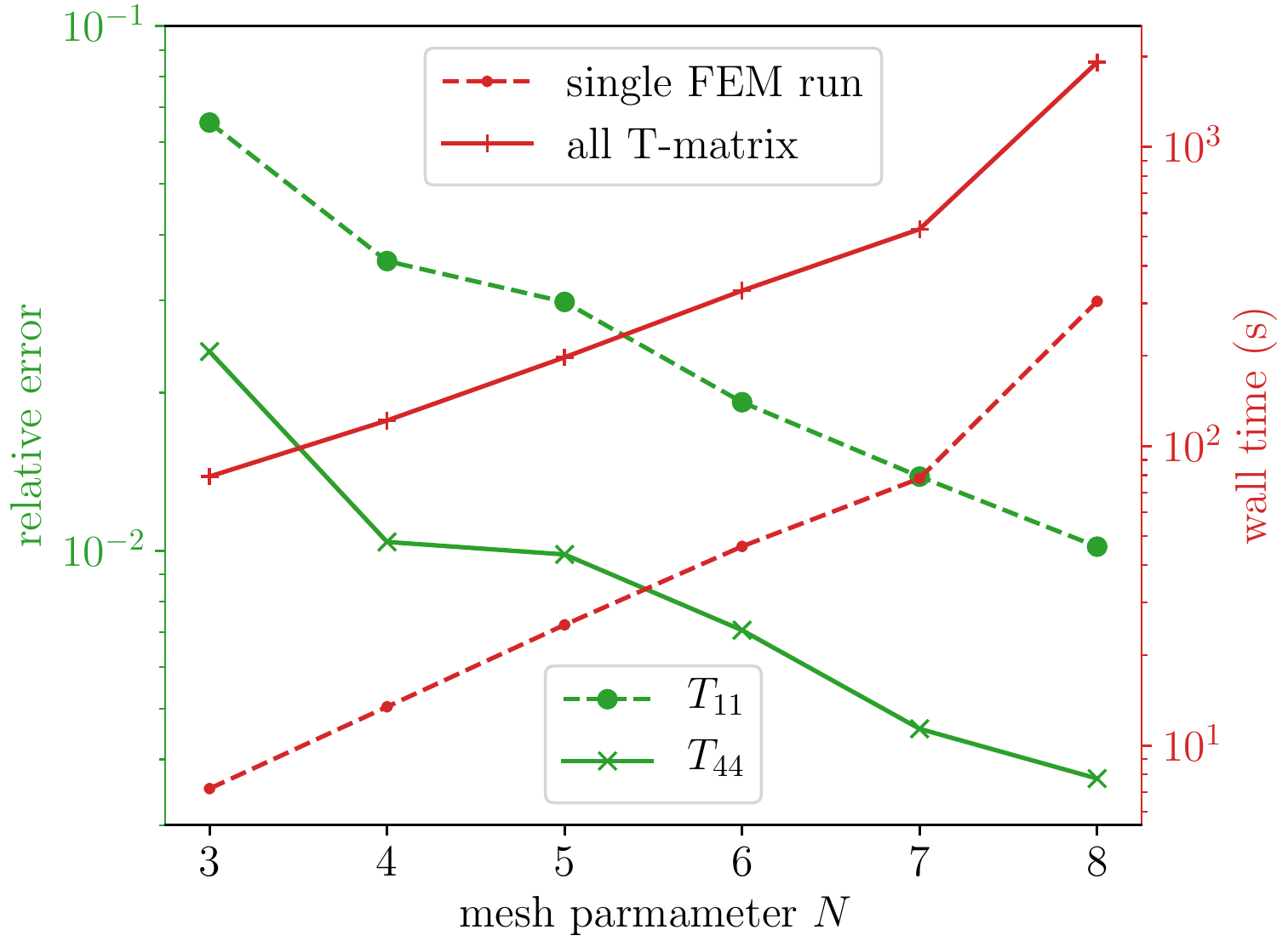}
		\caption{
      {\crevi Convergence and computation time for the sphere case as a function of the mesh refinement.
      The mesh refinement is parametrized by $N$, the number of mesh elements per wavelength inside the
      considered material (\emph{i.e.} the mesh size is set to $\lambda_0/(\sqrt{\varepsilon_r}N)$). The
      green dotted (resp. solid) line represents the evolution of the relative error
      $|T_{11}^{FEM}-T_{11}^{Mie}|/|T_{11}^{Mie}|$ (resp. $|T_{44}^{FEM}-T_{44}^{Mie}|/|T_{44}^{Mie}|$ ).
      The red dotted line represents the computation time for a single FEM run.
      The red solid  line represents the computation time of the full $T$-matrix.}
    } \label{fig:conv}
		\end{center}
\end{figure}

{\crevi The $T$-matrix of the sphere is deduced from the FEM computations for $n_{max}=1$ (\emph{i.e.} $p_{max}=3$). Indeed, in this
purely spherical configuration, the numerical FEM values of the $T$-matrix can
 be compared compared to Mie results. First, the proper
convergence of the dominant (diagonal) $T$-matrix coefficients are checked and shown in Fig.~(\ref{fig:conv}). This convergence test has
been performed on a laptop with 16Gb of memory with a 2.6GHz quad-core processor. The mesh refinement parametrized by $N$ defined above
ranges from 3 (coarse mesh leading to a sparse matrix of size 63945) to 8 (a moderate mesh leading to a sparse matrix of size 567227). The
green dotted (resp. solid) line represents the evolution of the relative error $|T_{11}^{FEM}-T_{11}^{Mie}|/|T_{11}^{Mie}|$ (resp.
$|T_{44}^{FEM}-T_{44}^{Mie}|/|T_{44}^{Mie}|$). The red dotted line represents the computation time for a single FEM run. The red solid
line represents the computation time for the full $T$-matrix. The number of FEM calculations is $2p_{max}$ to fill the full matrix. For
the fine mesh with $N=8$, one FEM run takes 5 minutes. The relative error with respect to the Mie coefficients is then close to $10^{-3}$ (resp.
$10^{-2}$) for electric (resp. magnetic) type $T$-matrix coefficients denoted here $f_{nm}^{(e)}$ (resp $f_{nm}^{(h)}$). Note that with
this type of dielectric contrast, the modulus of the electric type coefficients is expected to be higher that their magnetic
counterpart. This translates into the higher relative precision reached for the dominant coefficient. Interestingly enough, the relative
error with respect to Mie results for the dominant coefficient obtained with the very coarse mesh is less than $3\%$ with a runtime of
only 7s for one single FEM computation on a laptop. This can be valuable in optimization processes.

The geometrical order is set to 1 which means that we are dealing with planar tetrahedral elements. As a consequence, the sphere surface
is numerically tessellated (discretized by planar triangular elements). It is necessary to further refine the mesh on the sphere
boundary to retrieve Mie theory results more accurately. As a consequence, the mesh characteristic size is from now on set to
$\lambda_0/14$ (N=14) in the background and to one sixth of its value in free space inside the sphere. The numerical results presented in
the following were obtained on a desktop with 24 cores and 256Gb of RAM memory.}

For a sphere, the $T$-matrix is expected to be diagonal. The following theoretical considerations should hold and are
confronted with to the numerical results shown in Table~\ref{tab:sphere_T}:
\begin{enumerate} 
	\item \emph{The $T$-matrix should not depend on the radius $R$ considered for the expansion}:
				Table~\ref{tab:sphere_T} shows the real and imaginary parts
				(columns labelled $\RE\{T_{i,j}\}$ and $\IM\{T_{i,j}\}$) of the mean values 
				found for the radial cuts as well as their standard deviations $\sigma^{\mathrm{cuts}}_{i,j}$ 
				obtained for the diagonal elements of the $T$-matrix.
	\item \emph{Off-diagonal elements of the $T$-matrix should be null for a sphere}:
				The mean modulus of theoretically null matrix elements ($T_{i,j}$ with $i\neq j$) 
				is \num[round-mode=places, round-precision=1]{2.18927583828e-06},
				which is below the mean value of their standard deviation $\sigma^{\mathrm{cuts}}_{i,j}$ 
				(\num[round-mode=places, round-precision=1]{3.49691586771e-06}).
	\item \emph{Diagonal coefficients should not depend on the angular momentum $m$ for a sphere}:
				which is the case up to numerical precision (see \emph{e.g.} $T_{1,1}$, $T_{2,2}$ and $T_{3,3}$ 
				in Table~\ref{tab:sphere_T}).
	\item \emph{Coefficients should satisfy $|2\,T_{i,i}+1|=1$}, which is the case up to numerical 
				 precision as shown in the column labelled $|2\,T_{i,i}+1|$.
	\item The coefficients $f_{n,m}^{(e)}$ can be computed by making the use of the transverse
				components of the FEM calculated field only using Eq.~(\ref{eq:fenm_tangent}) or
				the normal component of the same field only using Eq.~(\ref{eq:fenm_normal}).
				They are nearly equal (see last lines of Table~\ref{tab:sphere_T}).
\end{enumerate}

The absolute precision when compared to Mie theory upon the $T$-matrix element is better than $10^{-3}$ for the mesh and PML parameters
chosen. Let us recall here that the discretized FEM sphere is tessellated because of the first order geometric order of the mesh.
This is the reason why the intrinsic error of the FEM method is lower ($\neq10^-5$) that the relative error with Mie coefficients ($\neq10^-3$).

\section{Results for an ellipsoid of revolution}
Finally, the present method applies to non-spherical objects. 
{\crevi In order to illustrate, we compare FEM results to those obtained with an in-house code based on the
Waterman $T$-matrix theory \cite{chew1995waves,auger2007mie,auger2007angularly}.}
We consider an spheroid with the following radii $r_x=r_y=\lambda/16$ and $r_z=\lambda/4$ with 
relative permittivity of 9 in a homogeneous background of relative permittivity of 1. 

\begin{figure}[h]
		\begin{center}
      \includegraphics[width=\columnwidth]{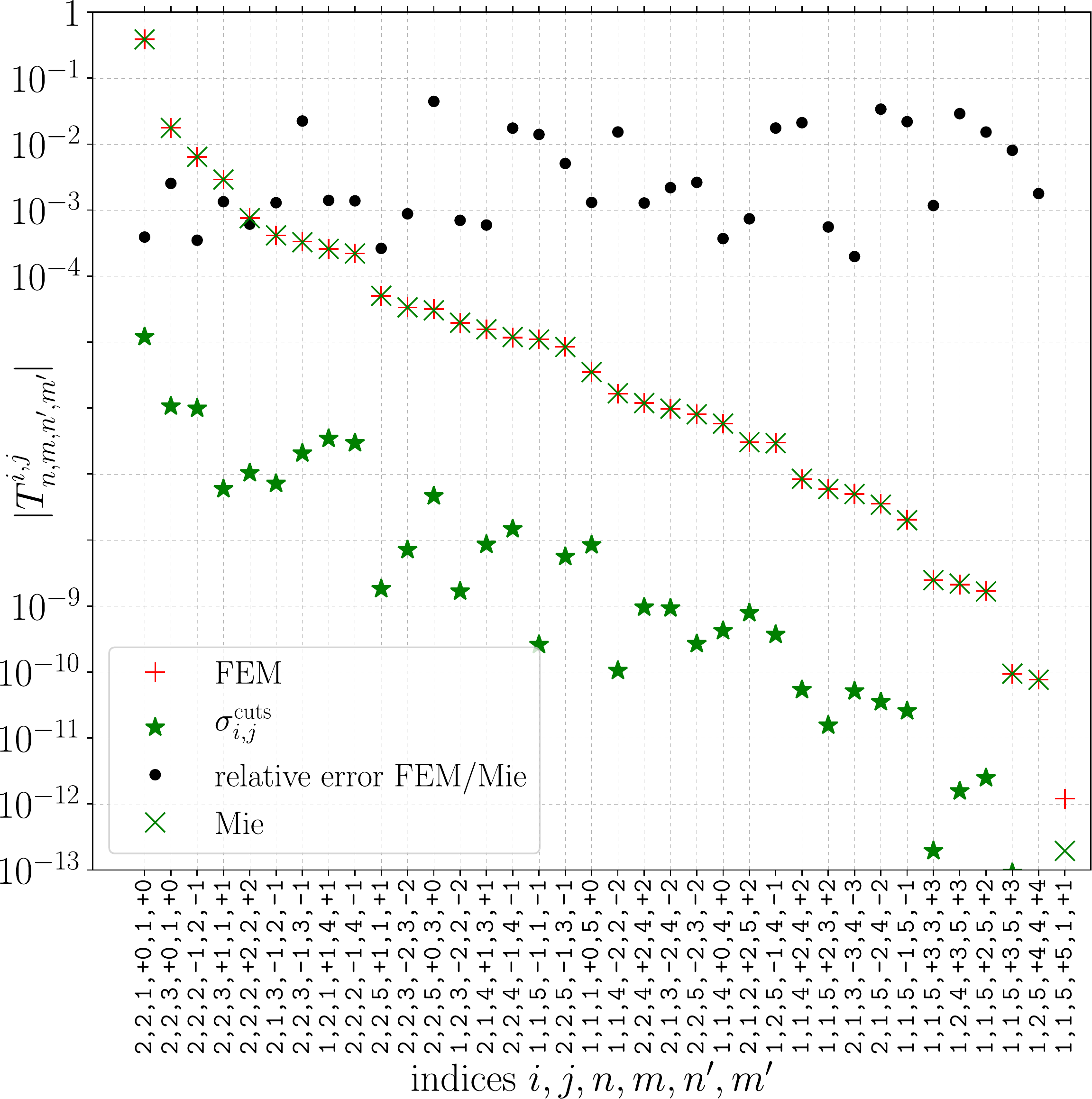}
		\caption{Coefficients of the $T$-matrix indexed by $i,j,n,m,n',m'$ for a spheroid. 
						 For clarity, only one over seven of the coefficients having a ratio mean value by
             standard deviation lower than 0.01 are represented, 
             sorted by decreasing modulus.} \label{fig:coef_ellips}
		\end{center}
\end{figure}

The numerical results obtained for the spheroid are shown in Fig.~\ref{fig:coef_ellips}.
with the two methods. Contrarily to the spherical case, the $T$-matrix is now 
expected to be non-diagonal and its dimensions are {\crevii $70\times70$} for $n_{max}=5$ 
(see Eq.~(\ref{eq:pmax})). Its coefficients were sorted in decreasing value of their modulus
and Fig.~\ref{fig:coef_ellips} shows one coefficient every seven having a ratio mean value 
by standard deviation over the 10 cuts lower than 0.01.
A striking observation from the figure is that coefficients with modulus as small as $10^{-10}$ 
can be retrieved. The relative error ($~10^{-2}$) compared to Waterman $T$-matrix is due to the 
intrinsic geometrical bias arising from the discretization of the spheroid by planar elements.

\section{Conclusion} 

We have shown that the $T$-matrix of an arbitrary scatterer can be retrieved very accurately using a diffracted field formulation of the
FEM. All the details necessary to its implementations are given explicitly. An open-source model allowing to retrieve the numerical
results is provided \cite{code}. Numerical results have been confronted to Generalized Mie theory with excellent agreement. At least 4
significant digits can be obtained using a $\lambda_0/14$ characteristic mesh parameter with second order shape functions. A
straightforward improvement of the method would indeed lie in the use of higher order curved elements to discretize the curved objects.
This type of hybrid method combining two rigorous and complementary numerical schemes allows to combine the strengths of both methods.
For instance, {\crevii it can be used to determine the total $T$-matrix of objects in close proximity by considering their global circumscribing
sphere,} or as an elementary brick of multiple scattering codes to tackle very large scattering systems made of arbitrarily shaped
particles. It also allows the fast determination of polarization or angular responses of a single scatterer.

\section{Appendix}
\subsection{Scalar spherical harmonics}
For positive $m$, the associated Legendre functions are given in most modern
texts and programs by
\begin{equation}
P_{n}^{m}(x)=(-1)^{m}\left(	 1-x^{2}\right)	 ^{m/2}\frac{d^{m}}{dx^{m}}%
P_{n}(x)
\end{equation}
for all $m=0,...,n$. With this convenient definition, scalar spherical
harmonics, $Y_{nm}(\theta,\phi)$, are simply proportional to the associated
Legendre functions $P_{n}^{m}(\cos\theta)$ multiplied by $e^{im\phi}$. We take
advantage of the proportionality to introduce convenient normalization
factors:%
\begin{align}
Y_{nm}\left(	\widehat{\mathbf{r}}\right)		&	 =Y_{nm}\left(	\theta
,\phi\right)	\equiv\left[	\frac{2n+1}{4\pi}\frac{(n-m)!}{(n+m)!}\right]
^{\frac{1}{2}}P_{n}^{m}\left(	 \cos\theta\right)	e^{im\phi}\nonumber\\
&	 =\gamma_{nm}\sqrt{n\left(	n+1\right)	}P_{n}^{m}\left(	\cos\theta\right)
e^{im\phi}\equiv\lambda_{nm}P_{n}^{m}\left(	 \cos\theta\right)	e^{im\phi
}\nonumber\\
&	 \equiv\overline{P}_{n}^{m}\left(	 \cos\theta\right)	\exp(im\phi)
\label{scalharm}%
\end{align}
where in the second line we have introduced the normalized Legendre functions,
$\overline{P}_{n}^{m}$ and introduced two normalization factors $\gamma_{nm}$
and $\lambda_{nm}$:%
\begin{align}
\gamma_{nm}&\equiv	\sqrt{\frac{\left(	2n+1\right)	 \left(	 n-m\right)	 !}{4\pi
n\left(	 n+1\right)	 \left(	 n+m\right)	 !}}\,,\\
\lambda_{nm}& \equiv\sqrt{\frac{\left(	2n+1\right)	 \left(	 n-m\right)	 !}{4\pi\left(
n+m\right)	!}}=\gamma_{nm}\sqrt{n\left(	n+1\right)	}\,. \label{gamnorm}%
\end{align}

The reason for these two definitions for normalization is that $\gamma_{nm}$
is a practical normalization factor for vector spherical harmonics, while
$\lambda_{nm}$ is more practical for scalar spherical harmonics. An advantage
of this normalization is that we never have to compute $\overline{P}_{n}^{m}(x)$ 
with negative values of $m$ since
\begin{equation}
\overline{P}_{n}^{-m}(x)=(-1)^{m}\overline{P}_{n}^{m}(x)\,.
\end{equation}
As long as the angles $\theta$ and $\phi$ are real valued variables, this
allows us to simply calculate the complex conjugate of $Y_{nm}\left(
\theta,\phi\right)	$:%
\begin{equation}
Y_{n,m}^{\ast}\left(	\theta,\phi\right)	=\left(	 -1\right)	^{m}%
Y_{n,-m}\left(	\theta,\phi\right)\,.
\end{equation}
Their parity properties are%
\begin{equation}
Y_{nm}\left(	-\widehat{\mathbf{r}}\right)	=Y_{nm}\left(	 \pi-\theta,\phi
+\pi\right)	 =\left(	-1\right)	 ^{n}Y_{nm}\left(	 \widehat{\mathbf{r}%
}\right)\,. \label{Sphpar}%
\end{equation}
The scalar spherical harmonics are normalized with respect to an integration
over the solid angles :
\begin{align}
&\int_{0}^{4\pi}d\Omega\,Y_{\nu\mu}^{\ast}(\theta,\phi)\cdot Y_{nm}(\theta ,\phi)	 \nonumber	\\
&\equiv\left(	 -1\right)	^{\mu}\int_{0}^{\pi}\sin\theta d\theta
\int_{0}^{2\pi}d\phi\,Y_{\nu,-\mu}(\theta,\phi)\cdot Y_{nm}(\theta
,\phi)\nonumber	 \\
& =\left(	 -1\right)	^{\mu}\int_{-1}^{1}d\left(	\cos\theta\right)	 \int
_{0}^{2\pi}d\phi\,Y_{\nu,-\mu}(\theta,\phi)\cdot Y_{nm}(\theta,\phi
)	 \nonumber \\
&=\delta_{n,\nu}\delta_{m,\mu} \label{ynorm} %
\end{align}

\subsection{Indexing spherical harmonics and Vector wave functions}\label{sec:indexing}

It is convenient to replace the double index $n=0,...,\infty$ and $m=-n,...,n$
by a single index $p$ defined by:%
\begin{equation}
p=n\left(	 n+1\right)	 -m \,.
\end{equation}
The inverse relations between a value of $p$ and the corresponding $n$, $m$
pair are given by:
\begin{align}
n\left(	 p\right)		&	 =\mathrm{Int}\left[	\sqrt{p}\right] \nonumber	 \,,\\
m\left(	 p\right)		&	 =n\left(	 p\right)	 \left[	 n\left(	p\right)
+1\right]	 -p	 \,.
\end{align}
One readily sees that the one-to-one correspondence between $p$ and a $n,m$
pair fills the following table:%
\begin{equation}%
\begin{array}
[c]{ccr}\hline 
p & n & m\ \\\hline
\ 0\	& \ 0\	&  0\	\\
\ 1\	& \ 1\	&  1\	\\
\ 2\	& \ 1\	&  0\	\\
\ 3\	& \ 1\	& -1\ \\
\ 4\	& \ 2\	&  2\	\\
\ 5\	& \ 2\	&  1\	\\
\ 6\	& \ 2\	&  0\	\\
\ 7\	& \ 2\	& -1\ \\
\ 8\	& \ 2\	& -2\ \\
\ \vdots\	 & \vdots\	& \vdots\ \ \\\hline
\end{array}
\end{equation} 

Thus for a maximum orbital number of $n_{\max}$, the number of elements in the
table is :
\begin{equation}\label{eq:pmax}
p_{\max}=\left[	 n_{\max}+1\right]	^{2}-1=n_{\max}^{2}+2n_{\max}	 \,.%
\end{equation}

\subsection{Recurrence relations for scalar spherical harmonics}

The normalized Legendre functions can be calculated via recurrence relations.
We determine some maximum order, $n_{\max}$ that we want to calculate. We
initialize the recurrence with:%
\begin{equation}
\overline{P}_{0}^{0}\left(	u\right)	=\sqrt{\frac{1}{4\pi}}	\,. \label{P0init}%
\end{equation}
We can then calculate all the $\overline{P}_{n}^{n}$ up to $n_{\max}$:
\begin{equation}
\overline{P}_{n}^{n}\left(	x\right)	=-\sqrt{\frac{2n+1}{2n}}\sqrt{1-x^{2}%
}\overline{P}_{n-1}^{n-1}\left(	 x\right)	 \qquad n=1,...,n_{\max}.
\label{nnrec}%
\end{equation}
The $\overline{P}_{n}^{m}$ with $m=n-1$ are calculated via the relations:%
\begin{equation}
\overline{P}_{n}^{n-1}\left(	x\right)	=x\sqrt{2n+1}\overline{P}_{n-1}%
^{n-1}\left(	x\right)	\qquad n=1,...,n_{\max}. \label{P0min1}%
\end{equation}
All the remaining $\overline{P}_{n}^{m}\left(	 u\right)	 $ with $m=1,...,n-2$
can be successively calculated for all $n=3,...,n_{\max}$ using the relations :%

\begin{align}
\overline{P}_{n}^{m}\left(	x\right)	&=\sqrt{\frac{2n+1}{n^{2}-m^{2}}} \nonumber \\ &\left[
\sqrt{\left(	2n-1\right)	 }x\overline{P}_{n-1}^{m}\left(	 x\right)
-\sqrt{\frac{\left[	 (n-1)^{2}-m^{2}\right]	 }{\left(	 2n-3\right)	}%
}\overline{P}_{n-2}^{m}\left(	 x\right)	 \right] \,.	\label{P0mn}%
\end{align}
All the $\overline{P}_{n}^{m}$ with negative values of $m$ are calculated
using :%
\begin{equation}
\overline{P}_{n}^{-m}\left(	 x\right)	 =\left(	-\right)	^{m}\overline{P}%
_{n}^{m}\left(	x\right) \,.
\end{equation}
This recurrence procedure is just the analogue for the one that used to
determine $\overline{u}_{n}^{m}$ functions for Vector Spherical Harmonics. In
practice, one generally doesn't need to calculate the $\overline{P}_{n}%
^{m}\left(	x\right)	$ with $\left\vert m\right\vert \neq0$, since these can
be directly obtained from the $\overline{u}_{n}^{m}\left(	 x\right)	 $ defined
in Eq.~(\ref{nuf}) for the vector spherical harmonics:%
\begin{equation}
\overline{P}_{n}^{m}(x)=\sqrt{1-x^{2}}\frac{\sqrt{n(n+1)}}{m}\overline{u}_{n}^{m}(x) \,.
\end{equation}
Unfortunately, this relation doesn't yield the $m=0$ elements of $\overline
{P}_{n}^{m}\left(	 x\right)	 $, but since we have the elements $\overline
{P}_{0}^{0}$ and $\overline{P}_{1}^{0}$, from Eq.~(\ref{P0init}) and
Eq.~(\ref{P0min1}) we can calculate all the other $\overline{P}_{n}^{0}\left(
u\right)	$ using the relation of Eq.~(\ref{P0mn}) restricted to the $m=0$
case:
\begin{align}
\overline{P}_{n}^{0}\left(	x\right)	&=\frac{\sqrt{2n+1}}{n} \nonumber\\
&\left[\sqrt{2n-1}\,x\overline{P}_{n-1}^{0}\left(	 x\right)	 -\frac{n-1}{\sqrt{2n-3}%
}\overline{P}_{n-2}^{0}\left(	 x\right)	 \right]\, \nonumber \\
& \mathrm{with}\, \,n=2,...,n_{\max} \,.
\end{align}

\subsection{Vector spherical harmonics}

The Vector Spherical Harmonics (VSHs), $\mathbf{X}_{nm}$, $\mathbf{Y}_{nm}$,
and $\mathbf{Z}_{nm}$ have the numerically convenient expressions :
\begin{align}
\mathbf{Y}_{nm}(\theta,\phi)	&	 =\overline{P}_{n}^{m}(\cos\theta)\exp(im\phi)\widehat{\mathbf{r}}\\
&=\gamma_{nm}\sqrt{n\left(	n+1\right)	}P_{n}%
^{m}\left(	\cos\theta\right)	 e^{im\phi}\widehat{\mathbf{r}}\label{Ynorm}\\
\mathbf{X}_{nm}(\theta,\phi)	&	 =i\overline{u}_{n}^{m}(\cos\theta)\exp
(im\phi)\widehat{\bm{\theta}}-\overline{s}_{n}^{m}(\cos\theta)\exp
(im\phi)\widehat{\bm{\phi}}\label{Xenuets}\\
\mathbf{Z}_{nm}(\theta,\phi)	&	 =\overline{s}_{n}^{m}(\cos\theta)\exp
(im\phi)\widehat{\bm{\theta}}+i\overline{u}_{n}^{m}(\cos\theta)\exp
(im\phi)\widehat{\bm{\phi}} \label{Zenuets}%
\end{align}
where one should remark that the $\mathbf{Y}_{nm}(\theta,\phi)$ are defined
for $n=0,...,\infty$, and $m=-n,...,n$. The transverse VSHs, $\mathbf{X}_{nm}%
$, and $\mathbf{Z}_{nm}$ are defined only starting with $n=1$.

We also remark that the transverse spherical harmonics are conveniently
expressed in terms of $\overline{u}_{n}^{m}$ and $\overline{s}_{n}^{m}$ which
are defined by~:
\begin{align}
\overline{u}_{n}^{m}(\cos\theta)	&	 \equiv\frac{1}{\sqrt{n(n+1)}}\frac
{m}{\sin\theta}\overline{P}_{n}^{m}(\cos\theta)\\&=\gamma_{nm}\frac{m}{\sin
\theta}P_{n}^{m}(\cos\theta)\label{nuf}\\
\overline{s}_{n}^{m}(\cos\theta)	&	 \equiv\frac{1}{\sqrt{n(n+1)}}\frac
{d}{d\theta}\overline{P}_{n}^{m}(\cos\theta)\\&=\gamma_{nm}\frac{d}{d\theta}%
P_{n}^{m}(\cos\theta) \label{nsf}%
\end{align}
Integrating the normalization into the definition of $\overline{u}_{n}^{m}$
and $\overline{s}_{n}^{m}$ one then completely avoids calculating factorial
functions. We remark in passing that vector products of the transverse VSHs
$\mathbf{X}_{nm}$ and $\mathbf{Z}_{nm}$ with $\widehat{\mathbf{r}}$, transform
from one into the other, that is:%
\begin{align}
\mathbf{X}_{nm}(\theta,\phi)	&	 =-\widehat{\mathbf{r}}\times\mathbf{Z}%
_{nm}(\theta,\phi)=\mathbf{Z}_{nm}(\theta,\phi)\times\widehat{\mathbf{r}%
}\nonumber\\
\mathbf{Z}_{nm}(\theta,\phi)	&	 =\widehat{\mathbf{r}}\times\mathbf{X}%
_{nm}(\theta,\phi)=-\mathbf{X}_{nm}(\theta,\phi)\times\widehat{\mathbf{r}}%
\end{align}
The vector spherical harmonics, $\mathbf{Y,}$ $\mathbf{X,}$ and $\mathbf{Z}$
are defined so as to be orthonormalized :
\begin{equation}
\int_{0}^{4\pi}d\Omega\mathbf{A}_{n^{\prime}m^{\prime}}^{\ast}(\theta
,\phi)\mathbf{\cdot B}_{nm}(\theta,\phi)=\delta_{nn^{\prime}}\delta
_{mm^{\prime}}\delta_{AB}	 \,, \label{VSHnorm}%
\end{equation}
with $\mathbf{A=Y,}$ $\mathbf{X,}$ or\textbf{\ }$\mathbf{Z}$ and
$\mathbf{B=Y,}$ $\mathbf{X,}$ or\textbf{\ }$\mathbf{Z}$.

\subsection{Recurrence relations for the \textit{u} and \textit{s} functions}

We initialize the recurrence of the $\overline{u}_{n}^{m}$ functions with :%
\begin{equation}
\overline{u}_{n}^{0}(\cos\theta)=0
\end{equation}
for all $n$, and
\begin{equation}
\overline{u}_{1}^{1}(\cos\theta)=-\frac{1}{4}\sqrt{\frac{3}{\pi}}%
\end{equation}
One can then obtain all the $\overline{u}_{n}^{n}(x)$ up to $n_{\max}$ with%
\begin{equation}
\overline{u}_{n}^{n}(x)=-\sqrt{\frac{n\left(	2n+1\right)	 }{2\left(
n+1\right)	\left(	n-1\right)	}}\sqrt{1-x^{2}}\overline{u}_{n-1}^{n-1}(x)	 \,.
\end{equation}
The $\overline{u}_{n}^{m}$ with $m=n-1$ are calculated via the relations%
\begin{equation}
\overline{u}_{n}^{n-1}(x)=\sqrt{\frac{\left(	2n+1\right)	 \left(	 n-1\right)
}{\left(	n+1\right)	}}x\overline{u}_{n-1}^{n-1}(x)	\,.
\end{equation}
All the remaining $\overline{u}_{n}^{m}\left(	 u\right)	 $ with $m=1,...,n-2$
can be successively calculated for all $n=3,...,n_{\max}$ using the relations :%

\begin{align}
\overline{u}_{n}^{m}(x)	 &	=\sqrt{\frac{\left(	 n-1\right)	 \left(
4n^{2}-1\right)	 }{\left(	 n+1\right)	 \left(	 n^{2}-m^{2}\right)	 }%
}x\,\overline{u}_{n-1}^{m}(x)\nonumber\\
&	 \hspace{.1cm}-\sqrt{\frac{\left(	 2n+1\right)	\left(	n-1\right)	\left(
n-2\right)	\left(	n-m-1\right)	\left(	n+m-1\right)	}{\left(	2n-3\right)
n\left(	 n+1\right)	 \left(	 n^{2}-m^{2}\right)	 }}\overline{u}_{n-2}^{m}(x)
\end{align}
or more compactly as
\begin{align}
\overline{u}_{n}^{m}(x) &=\sqrt{\frac{\left(	2n+1\right)	 \left(	 n-1\right)
}{\left(	n+1\right)	\left(	n^{2}-m^{2}\right)	}}\left[	x\sqrt{\left(
2n-1\right)	 }\,\overline{u}_{n-1}^{m}(x)\right. \nonumber \\&\hspace{.3cm}\left. -\sqrt{\frac{\left(	 n-2\right)
\left[	\left(	n-1\right)	^{2}-m^{2}\right]	 }{n\left(	2n-3\right)	 }%
}\overline{u}_{n-2}^{m}(x)\right]	 \,.
\end{align}
At the end of tis procedure one has all the non-negative $\overline{u}_{n}%
^{m}$ functions up to $n_{\max}$. The $\overline{u}_{n}^{m}(x)$ with negative
$m$ are simply obtained from:
\begin{equation}
\overline{u}_{n}^{-m}(x)=\left(	 -1\right)	^{m+1}\overline{u}_{n}^{m}(x)	 \,.
\end{equation}

Although one could have found the $\overline{s}_{n}^{m}$ functions using a
similar procedure, this is not necessary since one can readily tabulate the
$\overline{s}_{n}^{m}$ with positive $m$ using the recurrence relation:
\begin{align}
	\overline{s}_{n}^{m}(x)&=\frac{1}{m+1}\sqrt{\left(	n+m+1\right)	\left(
	n-m\right)	}\sqrt{1-x^{2}}\overline{u}_{n}^{m+1}(x) \nonumber
	\\&\hspace{2.6cm}+x\,\overline{u}_{n}^{m}(x)\,.
\end{align}
We find that the $\overline{s}_{n}^{m}$ with negative values of $m$ from the
property that,
\begin{equation}
\overline{s}_{n}^{-m}(x)=\left(	 -1\right)	^{m}\overline{s}_{n}^{m}(x) \,.
\end{equation}

\subsection{Recurrence relation for spherical Bessel functions}

The Ricatti Bessel, Neumann and Hankel functions are defined respectively:%
\begin{align}
\psi_{n}\left(	z\right)	&\equiv zj_{n}\left(	z\right)	,\\
\chi_{n}\left(	z\right)	&\equiv zy_{n}\left(	z\right)	,\\
\xi_{n}^{\left(+\right)	 }\left(	z\right) &\equiv zh_{n}^{\left(	 +\right)	 }\left(
z\right)	=\psi_{n}\left(	 z\right)	 +i\chi_{n}\left(	 z\right)\,.
\end{align}
We define $\varphi^{\left(	1\right)	}$, $\varphi^{\left(	2\right)	}$,
$\varphi^{\left(	+\right)	}$\ as the argument $z$ multiplying a `logarithmic
derivatives' of the Ricatti Bessel function, specifically:%
\begin{equation}
\varphi_{n}^{\left(	 1\right)	 }=z\frac{\psi_{n}^{\prime}\left(	 z\right)
}{\psi_{n}\left(	z\right)	},\quad\varphi_{n}^{\left(	2\right)	}%
=z\frac{\chi_{n}^{\prime}\left(	 z\right)	 }{\chi_{n}\left(	 z\right)	 }%
,\quad\varphi_{n}^{\left(	 3\right)	 }=z\frac{\xi_{n}^{\prime}\left(
z\right)	}{\xi_{n}\left(	 z\right)	 },%
\end{equation}
or more simply expressed:%
\begin{equation}
\varphi_{n}^{\left(	 1\right)	 }\left(	z\right)	\equiv\frac{\psi_{n}%
^{\prime}\left(	 z\right)	 }{j_{n}\left(	z\right)	},\quad\varphi
_{n}^{\left(	2\right)	}\left(	 z\right)	 \equiv\frac{\chi_{n}^{\prime
}\left(	 z\right)	 }{y_{n}\left(	z\right)	}\text{,}\quad\varphi
_{n}^{\left(	3\right)	}\left(	 z\right)	 \equiv\frac{\xi_{n}^{\prime}\left(
z\right)	}{h_{n}\left(	 z\right)	 }\,.
\end{equation}
The first few $\xi_{n}\left(	z\right)	$ functions are:
\begin{align}
\xi_{0}\left(	 z\right)		&	 =-ie^{iz}\nonumber\,,\\
\xi_{1}\left(	 z\right)		&	 =-e^{iz}\left(	 1+\frac{i}{z}\right) \,,	 \nonumber\\
\xi_{2}\left(	 z\right)		&	 =e^{iz}\left(	i-\frac{3}{z}-\frac{3i}{z^{2}
}\right)\,.
\end{align}
The regular Ricatti Bessel functions are:
\begin{equation}
\psi_{n}\left(	z\right)	=zj_{n}\left(	 z\right)\,,
\end{equation}
and the first few values are :%
\begin{align}
\psi_{0}\left(	z\right)	 &	=\sin z\nonumber\,,\\
\psi_{1}\left(	z\right)	 &	=\frac{\sin z}{z}-\cos z\,,\\
\psi_{2}\left(	z\right)	 &	=\left(	 \frac{3}{z^{2}}-1\right)	 \sin
z-\frac{3}{z}\cos z\,.
\end{align}
The first few Ricatti Neumann functions, $\chi_{n}\left(	z\right)	$, are%
\begin{align}
\chi_{0}\left(	z\right)	 &	=-\cos z\,,\nonumber\\
\chi_{1}\left(	z\right)	 &	=-\frac{\cos z}{z}-\sin z\,,\nonumber\\
\chi_{2}\left(	z\right)	 &	=-\left(	\frac{3}{z^{2}}-1\right)	\cos
z-\frac{3}{z}\sin z\,.
\end{align}
We can calculate the $\varphi_{n}^{\left(	 2\right)	 }$ from the upward
recurrence relation:%
\begin{equation}
\varphi_{n}^{\left(	 2\right)	 }\left(	z\right)	=\frac{z^{2}}{n-\varphi
_{n-1}^{\left(	2\right)	}\left(	 z\right)	 }-n\,,
\end{equation}
with an initialization of%
\begin{equation}
\varphi_{0}^{\left(	 2\right)	 }\left(	z\right)	=-z\frac{\sin z}{\cos z}\,.
\end{equation}
Once the $\varphi_{n}^{\left(	 2\right)	 }$ functions have been calculated,
one can readily generate the $\chi_{n}\left(	z\right)	$ functions with the
upward recurrence relation:%
\begin{equation}
\chi_{n}\left(	z\right)	=\frac{\chi_{n-1}\left(	 z\right)	 }{z}\left(
n-\varphi_{n-1}^{\left(	 2\right)	 }\left(	z\right)	\right)	 \label{neurec}%
\end{equation}
with an initialization of%
\begin{equation}
\chi_{0}\left(	z\right)	=-\cos z\,.
\end{equation}

The regular $\varphi_{n}\left(	z\right)	$ functions obey the same recurrence
relations as the $\varphi_{n}^{\left(	 2\right)	 }\left(	z\right)	$
functions. If one starts calculating them by upward recurrence, everything
usually works fine at the beginning, but at some value of $n$, the
recurrence relation goes completely off course and the values are completely
wrong from there on out. We follow the usual Bohren and Huffman
suggestion that the $\varphi_{n}\left(	z\right)	$ functions be calculated
starting from high values of $n$ in the reverse recurrence relation. Starting
with $n$ equal to at least $n_{\max}+20$ where $n_{\max}$ is the
largest value used in calculations with simply $\varphi_{n_{\max}+20}\left(	z\right)=0$ 
is usually a safe choice. The $\varphi_{n}\left(z\right)$ functions so 
obtained have always been the correct ones up to machine
precision. The reverse recurrence relation is:
\begin{equation}
\varphi_{n}\left(	 z\right)	 =n+1-\frac{z^{2}}{n+1+\varphi_{n+1}\left(
z\right)	} \,.\label{QBessel}
\end{equation}
One can check calculations by verifying that the $\varphi_{0}\left(	 z\right)
$ obtained by backward recurrence is equal to the analytical result:
\begin{equation}
\varphi_{0}\left(	 z\right)	 =z\frac{\cos z}{\sin z}\,.
\end{equation}

Once the $\varphi_{n}\left(	 z\right)	 $ functions have been calculated, one
can readily generate the $\psi_{n}\left(	z\right)	$ functions with the
upward recurrence relation:%
\begin{equation}
\psi_{n}\left(	z\right)	=\frac{\psi_{n-1}\left(	 z\right)	 }{z}\left(
n-\varphi_{n-1}\left(	 z\right)	 \right)	\,, \label{psirec}%
\end{equation}
starting with the initial value $\psi_{0}\left(	 z\right)	 =\sin z$.%
\begin{equation}
j_{n}\left(	 z\right)	 =\frac{j_{n-1}\left(	 z\right)	 }{z}\left(
n-\varphi_{n-1}\left(	 z\right)	 \right)\,.
\end{equation}

\subsection{Determining the field coefficients of vector partial wave expansions}\label{sec:pwe}

Vector partial waves (VPWs), also called Vector spherical waves are simple to
write in terms of the vector spherical harmonics. The regular transverse
waves, are traditionally denoted $\mathbf{M}_{nm}^{\left(	 1\right)	 }$ and
$\mathbf{N}_{nm}^{\left(	1\right)	}$. An individual $\mathbf{M}%
_{nm}^{\left(	 1\right)	 }$ or $\mathbf{N}_{nm}^{\left(	 1\right)	 }$ is a
electromagnetic mode and should be viewed as a stationary wave. One needs a
superposition of more than one $\mathbf{M}_{nm}^{\left(	 1\right)	 }$ and/or
$\mathbf{N}_{nm}^{\left(	1\right)	}$ mode to describe a propagating wave. In
fact, \textbf{any}\ propagating incident field can be described as a
superposition of $\mathbf{M}_{nm}^{\left(	 1\right)	 }$ and/or $\mathbf{N}%
_{nm}^{\left(	 1\right)	 }$ modes. Although $\mathbf{M}_{nm}^{\left(
1\right)	}$ and $\mathbf{N}_{nm}^{\left(	 1\right)	 }$ are orthogonal, they
have an infinite normalization when integrated over all space:
\begin{align}
\mathbf{M}_{nm}^{\left(	 1\right)	 }(k\mathbf{r})	 &	\equiv j_{n}\left(
kr\right)	 \mathbf{X}_{nm}(\theta,\phi)\\
\mathbf{N}_{nm}^{\left(	 1\right)	 }(k\mathbf{r})	 &	\equiv\frac{1}%
{kr}\left[	\sqrt{n\left(	 n+1\right)	 }j_{n}\left(	 kr\right)	\mathbf{Y}%
_{nm}(\theta,\phi)\right. \nonumber \\&\hspace{2.6cm}\left.+\psi_{n}^{\prime}\left(	 kr\right)	\mathbf{Z}_{nm}%
(\theta,\phi)\right]	\label{RgMetN}
\end{align}
where $\psi_{n}\left(	 x\right)	 $ is the Ricatti Bessel function:%
\begin{equation}
\psi_{n}\left(	x\right)	\equiv xj_{n}\left(	 x\right)\,.
\end{equation}
We don't even have to calculate the derivative since $\psi_{n}^{\prime}\left(
kr\right)	 $ is readily obtained from a recurrence relation:%
\begin{equation}
\psi_{n}^{\prime}\left(	 z\right)	 =\psi_{n-1}\left(	z\right)	-nj_{n}\left(
z\right)\,.
\end{equation}

Any field with outgoing boundary condition can be described on the basis of
outgoing VPWs:
\begin{align}\label{eq:defMN}
\mathbf{M}_{nm}^{\left(	 +\right)	 }(k\mathbf{r})	 &	\equiv h_{n}^{\left(
+\right)	}\left(	 kr\right)	\mathbf{X}_{nm}(\theta,\phi)\\
\mathbf{N}_{nm}^{\left(	 +\right)	 }(k\mathbf{r})	 &	\equiv\frac{1}%
{kr}\left[	\sqrt{n\left(	 n+1\right)	 }h_{n}^{\left(	 +\right)	 }\left(
kr\right)	 \mathbf{Y}_{nm}(\theta,\phi)\right.\nonumber\\&\hspace{3cm}\left.+\xi_{n}^{\prime}\left(	 kr\right)
\mathbf{Z}_{nm}(\theta,\phi)\right]
\end{align}
where $\xi_{n}$ is the Ricatti Hankel function :%
\begin{equation}
\xi_{n}\left(	 x\right)	 \equiv xh_{n}^{\left(	+\right)	}\left(	 x\right)\,.
\end{equation}
Note that the value of $\xi_{n}^{\prime}\left(	k\right)	$ is readily determined from
the recurrence relation:%
\begin{equation}
\xi_{n}^{\prime}\left(	z\right)	=\xi_{n-1}\left(	z\right)	-nh_{n}^{\left(
+\right)	}\left(	 z\right)\,.
\end{equation}

These functions are also orthonormal with infinite overlap, but unlike the
$\mathbf{M}_{nm}^{\left(	1\right)	}$ and $\mathbf{N}_{nm}^{\left(	 1\right)
}$ they have an essential singularity at the origin.
Any scattered or \textquotedblleft outgoing" field can be expanded as:%
\begin{equation}
\mathbf{E}_{\mathrm{scat}}\left(	\mathbf{r}\right)	 =\sum_{n=1}^{n_{\max}%
}\sum_{m=-n}^{m=n}\left[	\mathbf{M}_{nm}^{\left(	 +\right)	 }(k\mathbf{r}%
)f_{nm}^{\left(	 h\right)	 }+\,\mathbf{N}_{nm}^{\left(	+\right)
}(k\mathbf{r})f_{nm}^{\left(	e\right)	}\right]\,.
\end{equation}
We can find its coefficients $f_{nm}^{\left(	h\right)	}$ and $f_{nm}^{\left(	e\right)	}$
by integrating on a sphere of any radius $R$ containing the scatterer:
\begin{align}\label{eq:fenm_normal}
f_{nm}^{\left(	e,\perp \right)	 }	&	 =\frac{kR}{h_{n}^{\left(	 +\right)	 }\left(
kR\right)	 \sqrt{n\left(	n+1\right)	}}\int_{0}^{4\pi}d\Omega\,\mathbf{E}%
_{\mathrm{scat}}\left(	R\widehat{\mathbf{r}}\right)	\cdot\mathbf{Y}%
_{nm}^{\ast}(\theta,\phi)\nonumber\\
&	 =\frac{kR}{h_{n}^{\left(	 +\right)	 }\left(	kR\right)	 \sqrt{n\left(
n+1\right)	}}\int_{0}^{4\pi}d\Omega\,E_{r}\left(	 R,\theta,\phi\right)
\overline{P}_{n}^{m}\left(	\cos\theta\right)	 e^{-im\phi}\,.
\end{align}
A possible check on this calculation is to calculate the $f_{nm}^{\left(
e\right)	}$ coefficients from the transverse components as well:
\begin{align}\label{eq:fenm_tangent}
f_{nm}^{\left(	e,\parallel\right)	}	 &	=\frac{kR}{\xi_{n}^{\prime}\left(	 kR\right)
}\int_{0}^{4\pi}d\Omega\,\mathbf{E}_{\mathrm{scat}}\left(	 R\widehat
{\mathbf{r}}\right)	 \cdot\mathbf{Z}_{nm}^{\ast}(\theta,\phi)\nonumber\\
&	 =\frac{kR}{\xi_{n}^{\prime}\left(	kR\right)	 }\int_{0}^{4\pi}%
d\Omega\left[	 E_{\theta}\left(	 R,\theta,\phi\right)	 \overline{s}_{n}%
^{m}(\cos\theta)\right.\nonumber\\ & \left. \hspace{2cm} -iE_{\phi}\left(	 R,\theta,\phi\right)	 \overline{u}_{n}%
^{m}(\cos\theta)\right]	 e^{-im\phi}\,.
\end{align}
In an analogous fashion, one can obtain the $f_{nm}^{\left(	 h\right)	 }$
coefficients from the transverse components:%
\begin{align}
f_{nm}^{\left(	h\right)	}	 &	=\frac{1}{h_{n}^{\left(	 +\right)	 }\left(
kR\right)	 }\int_{0}^{4\pi}d\Omega\,\mathbf{E}_{\mathrm{scat}}\left(
R\widehat{\mathbf{r}}\right)	\cdot\mathbf{X}_{nm}^{\ast}(\theta
,\phi)\nonumber\\
&	 =-\frac{1}{h_{n}^{\left(	 +\right)	 }\left(	kR\right)	 }\int_{0}^{4\pi
}d\Omega\left[	E_{\phi}\left(	R,\theta,\phi\right)	\overline{s}_{n}%
^{m}(\cos\theta)\right.\nonumber\\ & \left. \hspace{2cm} +iE_{\theta}\left(	 R,\theta,\phi\right)	 \overline{u}_{n}%
^{m}(\cos\theta)\right]	 e^{-im\phi}\,.
\end{align}

%
%
%
%

\bibliographystyle{plain}
\bibliography{biblio}

 
\end{document}